\documentclass[aps,prl,notitlepage,twocolumn]{revtex4-1}
\usepackage{hyperref}
\usepackage[x11names, rgb, html]{xcolor}
\definecolor{sbase03}{HTML}{002B36}
\definecolor{sbase02}{HTML}{073642}
\definecolor{sbase01}{HTML}{586E75}
\definecolor{sbase00}{HTML}{657B83}
\definecolor{sbase0}{HTML}{839496}
\definecolor{sbase1}{HTML}{93A1A1}
\definecolor{sbase2}{HTML}{EEE8D5}
\definecolor{sbase3}{HTML}{FDF6E3}
\definecolor{syellow}{HTML}{B58900}
\definecolor{sorange}{HTML}{CB4B16}
\definecolor{sred}{HTML}{DC322F}
\definecolor{smagenta}{HTML}{D33682}
\definecolor{sviolet}{HTML}{6C71C4}
\definecolor{sblue}{HTML}{268BD2}
\definecolor{scyan}{HTML}{2AA198}
\definecolor{sgreen}{HTML}{859900}
\hypersetup{
  colorlinks = true,
  citecolor = {sred},
  urlcolor = {scyan}
}
\usepackage{graphicx}  
\usepackage{dcolumn}   
\usepackage{bm}        
\usepackage{amssymb,amsmath}   
\usepackage[margin=1in]{geometry}

\newcommand{\avg}[1]{\langle #1 \rangle}

\begin{abstract}
We show that current fluctuations in a stochastic pump can be robustly mapped to fluctuations in a corresponding time-independent nonequilibrium steady state. 
We thus refine a recently proposed mapping so that it ensures equivalence of not only the averages, but also optimal representation of fluctuations in currents and density.
Our mapping leads to a natural decomposition of the entropy production in stochastic pumps similar to the ``housekeeping'' heat.
As a consequence of the decomposition of entropy production, the current fluctuations in weakly perturbed stochastic pumps are shown to satisfy a universal bound determined by the steady state entropy production.
\end{abstract}

\begin{document}
\title{Mapping current fluctuations of stochastic pumps to nonequilibrium steady states}

\author{Grant M. Rotskoff}
\email{rotskoff@berkeley.edu}
\affiliation{Biophysics Graduate Group, University of California, Berkeley, CA 94720, USA}

\date{\today}

\maketitle

Nonequilibrium steady states are an essential paradigm for describing nanoscale biological machines, such as molecular motors that extract work from chemical gradients~\cite{Yoshida2001}.
When a system is coupled to reservoirs with different chemical potentials, the dynamics breaks detailed balance and persistent, directed motion can be used to perform mechanical work.
Such a system is typically described as Markov processes with time-independent rates that depend both on the external chemical gradient and internal dynamics.

\begin{figure}[ht!]
\includegraphics[width=\linewidth]{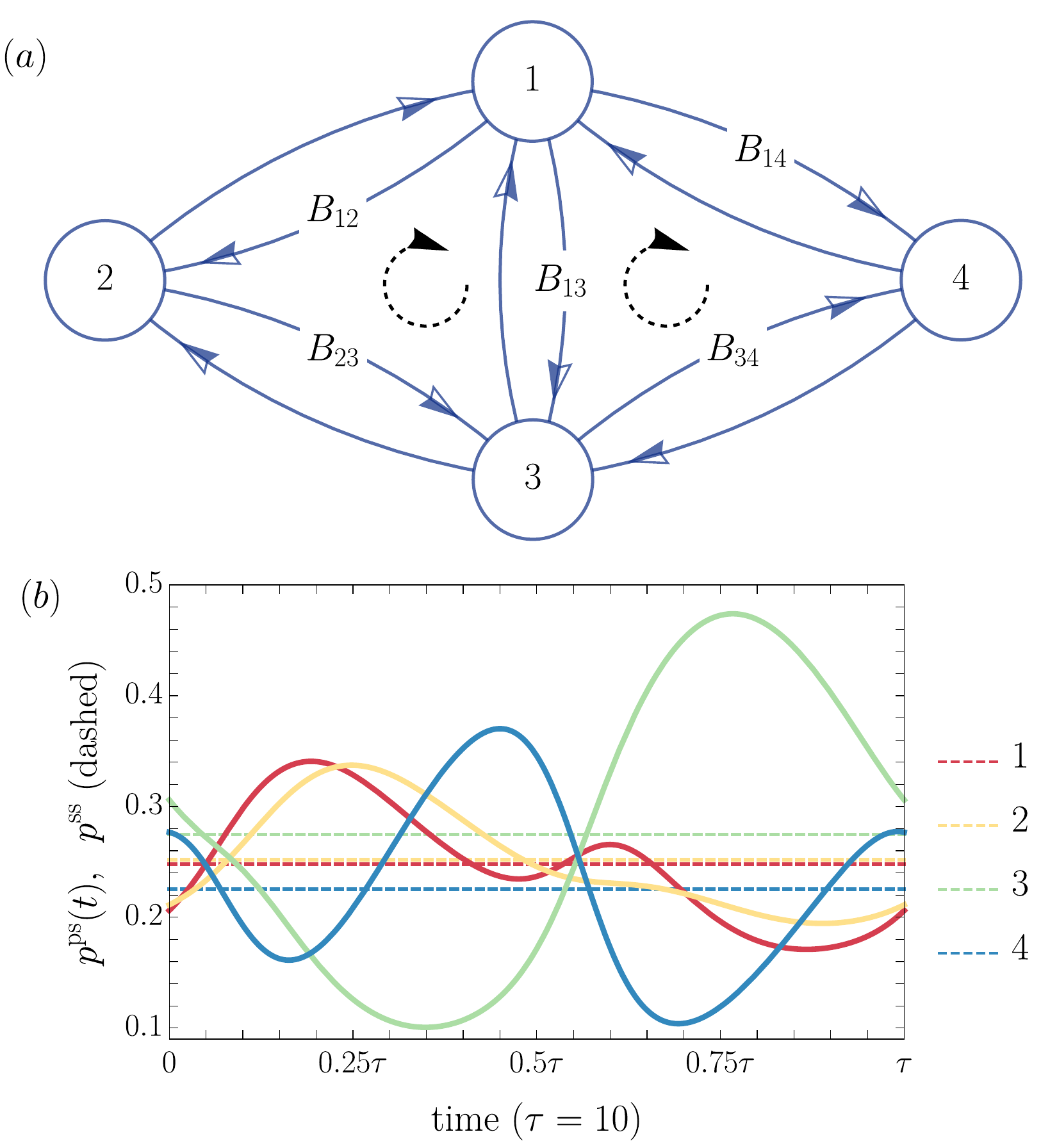}
  \caption{
  $(a)$ A schematic of the stochastic pump under consideration. 
  Symmetric barriers $B_{ij}$ and energy levels $E_i$ parametrize Arrhenius rates and are varied periodically in time to generate a current. 
  The corresponding nonequilibrium steady state representation of the pump has no time-dependence, but rather rates that break detailed balance.
  $(b)$ The time periodic steady state probabilities for each site on the graph are shown over an entire period $\tau$.
  The solid lines show the time-dependent occupation of the pump.
  The dashed lines show the average occupancy per period, a property matched by the corresponding steady state.  
  }
\label{fig:pump}
\end{figure}

Promising applications across many disciplines have motivated efforts to design artificial molecular machines that behave like those in biological settings.
Nonequilibrium steady states, however, have proved difficult to engineer~\cite{Wilson2016}.
Time-dependent external perturbations offer an alternative route to breaking detailed balance.
Indeed, many synthetic nanoscale machines are implemented as ``stochastic pumps,'' in which currents are generated by periodically varying an external potential~\cite{Browne2006,Sinitsyn2007,Horowitz2009,Sinitsyn2009,Martinez2016}.
A stochastic pump can be modeled as a non-homogeneous Markov jump process with instantaneous Arrhenius rates that are determined by time-dependent energy levels and barrier heights~\cite{Rahav2008, Mandal2011, Raz2016}.

Recently, \citet{Raz2016} proposed a mapping between time-independent steady states and periodically driven stochastic pumps that offers a set of design principles for engineering biomimetic nanodevices. 
While the mapping ensures that the average properties are asymptotically equivalent in both representations, it makes no guarantees about the fluctuations.
At the nanoscale, however, fluctuations play a crucial in determining characteristics like work and efficiency in finite-time measurements~\cite{Verley2014,Gingrich2014}.

Translating between nonequilibrium steady states and stochastic pumps relies on the so-called ``dynamical equivalence principle'' of Zia and Schmittmann~\cite{Zia2007}.
This principle stipulates that nonequilibrium steady states are characterized by the average currents and the average density. 
For Markov jump processes, the asymptotic fluctuations of a nonequilibrium steady state, however, are not dictated by these average properties alone.

Developments in large deviation theory, in particular the Level-2.5 formalism, have provided a general characterization of fluctuations away from the average behavior in Markov jump processes and diffusions~\cite{Maes2008,Bertini2015Large,Barato2015Formal}.
In this framework, both the average currents and their fluctuations are uniquely determined by the empirical density,
\begin{equation}
\rho_x = \frac{1}{t_{\rm obs}} \int_0^{t_{\rm obs}}dt\ \delta_{z(t),x},
\end{equation}
with $z(t)$ denoting the state at time $t$, and the empirical flow,
\begin{equation}
  q_{yx} = \frac{1}{t_{\rm obs}} \int_0^{t_{\rm obs}}dt\ \delta_{z(t^{-}),x} \delta_{z(t^+),y},
\end{equation}
which, roughly, counts the number hops from state $x$ to $y.$
It is important to note that the empirical flow contains more information than the empirical current; the latter specifies only the difference between the flow in the forward and reverse directions $j_{xy} = q_{xy} - q_{yx}.$
For example, the empirical current would not distinguish between a trajectory in which there are 100 $x \to y$ hops and $80$ $y \to x$ hops from one in which there are 20 $x \to y$ hops and $0$ in the opposite direction, while the flows would be dramatically different.
The large deviation rate function, $I(\rho,q),$ quantifies the rate of decay of probability of a joint observation of density and flow, 
\begin{equation}
P(\rho, q) \asymp \exp (-t_{\rm obs} I(\rho,q)).
\end{equation}
The symbol $\asymp$ indicates a logarithmic equivalence between $I(\rho,q)$ and $\lim_{t_{\rm obs}\to \infty} -1/t_{\rm obs} \ln P(\rho, q)$.
For both jump processes and diffusions, the rate function $I$ can be calculated explicitly~\cite{Maes2008,Bertini2015Large}.
Once the joint rate function for empirical density and empirical flow is known, fluctuations in currents can be computed via the contraction principle~\cite{Touchette2009}. 

The large deviation formalism suggests a stricter requirement for dynamical equivalence among jump processes: if the asymptotic form of the fluctuations is to be accurately captured, then it is not the average currents, but rather the average flows that must be used to describe the dynamics of a nonequilibrium steady state. 
This is a more rigid prescription, as detailed below.
Further, these insights motivate a solution to the mapping problem between stochastic pumps and nonequilibrium steady states that preserves the fluctuations.
Interestingly, in order to optimally describe current fluctuations of a stochastic pump, the corresponding nonequilibrium steady state must have a lower average entropy production rate than that of the pump.
The origin of this ``excess'' entropy production can be explained with a simple decomposition of the entropy production of the stochastic pump~\cite{Oono1998,Hatano2001,Esposito2008}.

The nonequilibrium steady state representation of the pump satisfies a universal lower bound on the magnitude of its current fluctuations, dictated by the total entropy production less the excess~\cite{Barato2015, Pietzonka2016, Gingrich2016}.
As a consequence of this splitting, we demonstrate that, in a perturbative limit, stochastic pumps satisfy a universal bound on their current fluctuations, dictated by the entropy production of the corresponding steady state. 
Taken together, these insights offer a powerful set of design principles for translating between stochastic pumps and steady states.

To illustrate our mapping, we consider a simple model of a stochastic pump: a single particle hopping with Arrhenius rates on a four state graph.
We vary one energy level and one barrier periodically in time, which is the minimal time-dependent perturbation that generates a non-vanishing current according to the no-pumping theorem~\cite{Sinitsyn2007,Sinitsyn2009,Horowitz2009}. 
This setup is depicted in Fig.~\ref{fig:pump} $(a)$.

The pump achieves a periodic steady state, which can be calculated numerically by integrating, 
\begin{equation}
  p_i^{\rm ps}(t+s) = \int_t^s W_{ij}(t+t') p_j^{\rm ps}(t+t') dt'.
\end{equation}
Here $p_i^{\rm ps}(t)$ is the probability of being in state $i$ at time $t$ and $W(t)$ is the continuous time rate matrix for the dynamics at time $t.$  
The periodic steady state satisfies
\begin{equation}
  p_i^{\rm ps}(t+\tau) = p_i^{\rm ps}(t),
\end{equation}
where $\tau$ is the period of the pumping protocol.
Note that, by construction, $W(t)$ satisfies detailed balance at each point in time. 
The Arrhenius rates determine the instantaneous rate matrix
\begin{align}
\nonumber  W_{ij}(t) = e^{-\beta \bigl(B_{ij}(t) - E_j(t) \bigr)} && \text{for } i\neq j \\
 \nonumber  W_{ii}(t) = -\sum_{i\neq j} W_{ij}(t)
\end{align}
where $E_j(t)$ denotes the energy level of state $j$ and $B_{ij}(t)=B_{ji}(t)$ is the barrier height.
In our example, the only time-dependent quantities are
\begin{equation}
\begin{aligned}
  E_3(t) &= \sin(2\pi t/\tau) + 1,\\ 
  E_4(t) &= \sin(4\pi t/\tau),\\
  B_{13}(t)&= \sin(2\pi t/\tau).
\end{aligned}
\end{equation}
The periodic solution is plotted in Fig.~\ref{fig:pump} $(c)$. 

We aim to find a time-independent rate matrix $W^{\rm ss}$ that mimics the stochastic pump and matches its fluctuations.
Following~\cite{Zia2007}, we let $\overline{W}_{ij} = {W}^{\rm ss}_{ij}\hat{p}_j$ where $\hat{p}_j$ is the average occupancy in the periodic steady state and write
\begin{equation}
  \overline{W}_{ij} = \mathcal{S}_{ij} + \mathcal{A}_{ij},
\end{equation}
where $\mathcal{S}$ is a symmetric, stochastic matrix and $\mathcal{A}$ is an antisymmetric matrix. 
The symmetric part of this decomposition is related to the ``activity'' of a trajectory~\cite{Lecomte2007,Baiesi2009}.
The continuous time rate matrix for the dynamics is then given by
\begin{equation}
  W^{\rm ss} = \left( \mathcal{S} + \mathcal{A} \right)\mathcal{P}^{-1}, 
\end{equation}
where $\mathcal{P}$ is a diagonal matrix with $\mathcal{P}_{ii} = \hat{p}_i,$ the steady state probability of site $i.$
If we further impose the constraint that the steady state currents agree with the periodic average current along each edge,
\begin{equation}
\hat{j}_{ij} = \int_0^{\tau} dt\  W_{ij}(t)p^{\rm ps}_j(t) - W_{ji}(t)p^{\rm ps}_i(t),
\end{equation}
then the antisymmetric part of the rate matrix is uniquely identified,
\begin{align}
  \mathcal{A}_{ij} = \frac{1}{2} \hat{j}_{ij}.
\end{align}
The rate matrix $W^{\rm ss}$ describes a probability conserving stochastic process, and, as a result, the form of $\mathcal{S}$ is constrained, but only weakly.
In particular, it must be the case that 
\begin{equation}
\mathcal{S}_{ij}\geq \left| \mathcal{A}_{ij}\right| 
\end{equation}
and
\begin{equation}
  \sum_j \mathcal{S}_{ij} = 0,
\end{equation}
which ensures that $W^{\rm ss}$ is a stochastic matrix.

Though the rate matrix is not uniquely specified, any valid choice of $\mathcal{S}$ results in a stochastic process with identical average currents and average occupancy statistics.
The same cannot be said for the fluctuations. 
The freedom in $\mathcal{S}$ can be directly represented by noting that any valid off-diagonal entry in the matrix can be written,
\begin{equation}
  \mathcal{S}_{ij} = c_{ij} | \mathcal{A}_{ij} |,\ c_{ij} > 1.
\end{equation}
Due to symmetry, there are $N(N-1)/2$ choices to make.
Indeed, the rate matrices resulting from different choices of $\mathcal{S}$ yield different average entropy production rates, given by,
\begin{equation}
  \hat{\sigma}_{ij} = \hat{j}_{ij}
    \ln \frac{c_{ij} |\hat{j}_{ij}| + \hat{j}_{ij} }
             {c_{ij} |\hat{j}_{ij}| - \hat{j}_{ij} }.
\end{equation}
The $c_{ij}$ values can be varied independently so long as they meet the constraint $c_{ij}\geq 1,$ meaning that the total entropy production can be made arbitrarily small by taking $c_{ij}$ large.

\begin{figure}[ht!]
\includegraphics[width=\linewidth]{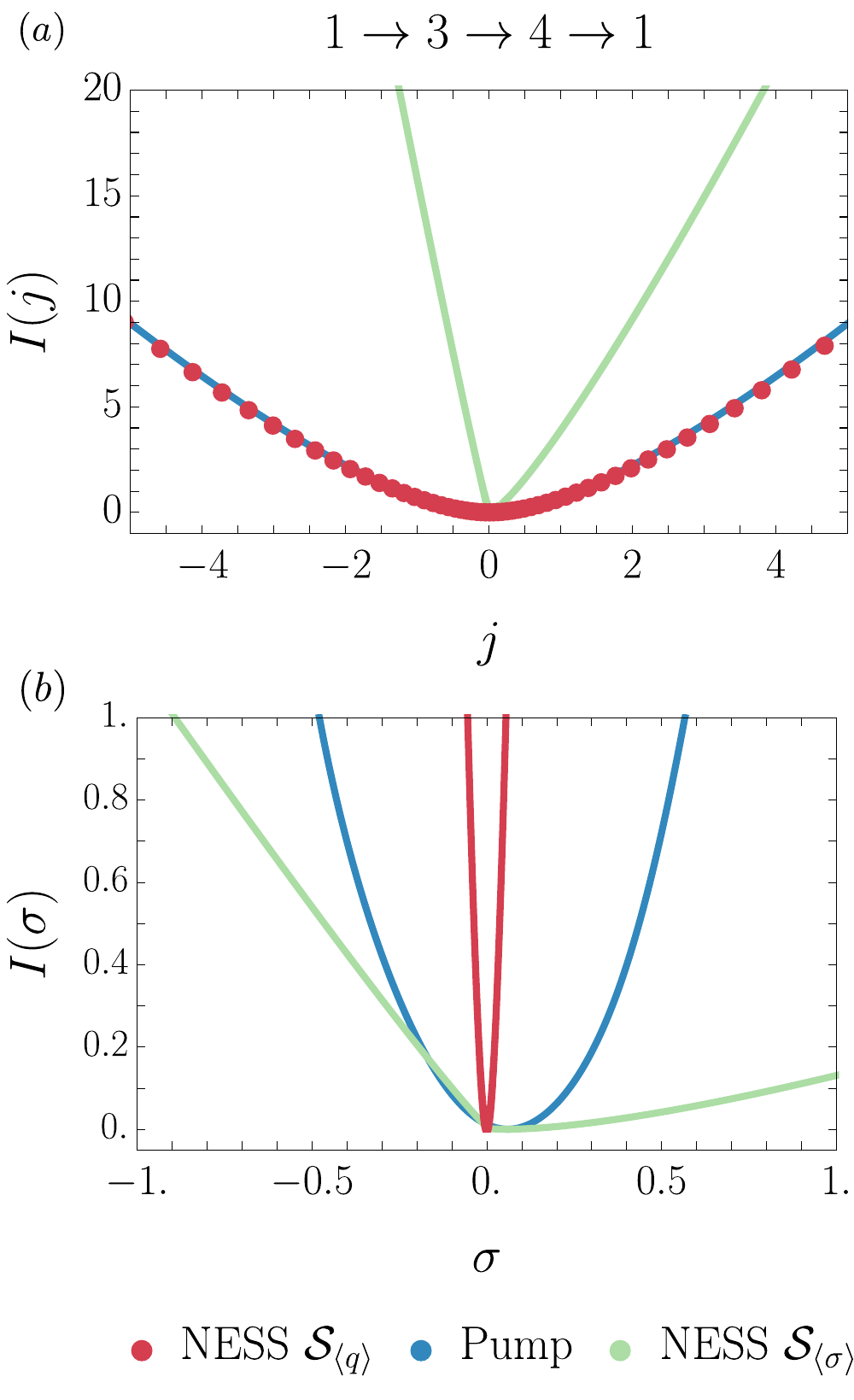}
\caption{
  $(a)$ The large deviation rate function for the current around the upper cycle (see Fig.~\ref{fig:pump}) $1\to 3\to 4\to 1$ is shown for the stochastic pump (blue) the nonequilibrium steady state with the same average entropy production along each edge as the pump (green), and the nonequilibrium steady state with the same average flow along each edge as the pump (red dots). 
  While the nonequilibrium steady state with $\mathcal{S}_{\avg{\sigma}}$ has the same average current, the character and extent of its fluctuations are extremely different. 
  Choosing $\mathcal{S}_{\avg{q}}$ preserves even very rare fluctuations in current.
  $(b)$ The large deviation rate functions for entropy production reveal that the steady state that recapitulates the current fluctuations has a smaller average entropy production.
  Furthermore, the extent of entropy production fluctuations in the corresponding steady state is much less pronounced. 
  $\mathcal{S}_{\avg{\sigma}}$, on the other hand, leads to greatly enhanced entropy production fluctuations.
}
\label{fig:rfs}
\end{figure}

\citet{Raz2016} suggest choosing $\mathcal{S}$ so that the average entropy production rate along each edge is the same in the stochastic pump and the nonequilibrium steady state representations.
This choice, which we denote $\mathcal{S}_{\avg{\sigma}}$ uniquely specifies a rate matrix and also guarantees that the average current, occupancy, and entropy production rates are preserved by the map.
However, the asymptotic fluctuations in entropy production and current are dramatically different.

To demonstrate this, we computed the entropy production and current large deviation rate functions for both the stochastic pump and the nonequilibrium steady state representation, shown in Fig.~\ref{fig:rfs}.
To calculate the rate functions, we first compute the scaled cumulant generating functions for entropy production $\omega$ and current $j$,
\begin{align}
  \psi_\omega(\lambda) = \lim_{t\to \infty} \frac{1}{t} \ln \avg{ e^{-\lambda \omega} }, &&
\psi_j (s) = \lim_{t\to \infty} \frac{1}{t} \ln \avg{ e^{-s j} }.
  \label{eq:cgfs}
\end{align}
For the nonequilibrium steady state representation, the cumulant generating functions can be calculated exactly by Cram\'{e}r tilting~\cite{Touchette2009}.
In the case of the stochastic pump, the averages in~\eqref{eq:cgfs} can be directly evaluated in the time-periodic steady state, meaning that the cumulant generating function can be numerically computed as,
\begin{equation}
  \psi_{\omega}(\lambda) = \frac{1}{\tau} \ln \sum_{j} \int_0^{\tau} W_{ij}(t;\lambda) p^{\rm ps}_j(t),
\end{equation}
where $W(t; \lambda)$ is the tilted rate matrix for entropy production~\cite{Lebowitz1999}.
We use the G\"{a}rtner-Ellis theorem to compute the large deviation rate functions by first computing the scaled cumulant generating function and then performing a Legendre-Fenchel transform~\cite{Touchette2009}.

Fig.~\ref{fig:rfs} $(b)$ shows the entropy production rate function with $W^{\rm ss} = \left( \mathcal{S}_{\avg{\sigma}} + \mathcal{A} \right) \mathcal{P}^{-1}$.
Note that, while the averages agree, the nature of the entropy production fluctuations is quite different. 
The steady state with the matching average entropy production has a notably fatter tail for large entropy production rates.

\emph{Excess entropy production.---}
In order to match the fluctuations in current, we instead choose $\mathcal{S}$ so that the average empirical flows are accurately captured by the network. 
In particular, we let 
\begin{equation}
  \mathcal{S}_{\avg{q}} = \hat{q}_{ij} - \frac{1}{2}\hat{j}_{ij} \iff \overline{W}_{ij} = \hat{q}_{ij},
  \label{eq:sflow}
\end{equation}
where $\hat{q}_{ij}$ denotes the average flow along edge $ij$ in the periodic steady state.
This choice has the additional advantage of simplicity: the dynamics produces the correct average number of hops in both directions along each edge of the network. 
We note that for high-dimensional networks, measuring all of the detailed edge currents or flows could be a formidable challenge. 
Because $\mathcal{S}$ does not affect the antisymmetric part of the rate matrix, the average currents along each edge are equivalent in both the stochastic pump and the nonequilibrium steady state. 
As illustrated by Fig.~\ref{fig:rfs} (a), choosing $\mathcal{S}_{\avg{q}}$ leads to striking agreement between the current fluctuations of the stochastic pump and the corresponding steady state. 

However, with the choice of $\mathcal{S}_{\avg{q}}$, both the average entropy production rate and its fluctuations in the nonequilibrium steady state representation differ markedly from the corresponding stochastic pump, as shown in Fig.~\ref{fig:rfs} $(b)$.
The ``excess'' entropy production has a physical origin and can be explained with a natural decomposition of the stochastic pump entropy production. 
Unlike nonequilibrium steady states, which can only produce entropy around closed cycles, stochastic pumps can produce entropy without completing a cycle~\cite{Esposito2010}.
We decompose the total stochastic pump entropy production rate into a contribution from the steady state, akin to the ``housekeeping heat'', and the excess associated with the pumping protocol~\cite{Oono1998,Hatano2001},
\begin{equation}
\sigma^{\rm pump} = \sigma^{\rm ss} + \sigma^{\rm ex},
\end{equation}
where, 
\begin{equation}
\sigma^{\rm ss}_{ij} = \hat{j}_{ij} 
                       \ln\frac{\hat{q}_{ij}}{\hat{q}_{ji}},
\end{equation}
and,
\begin{equation}
\sigma^{\rm ex}_{ij} = \frac{1}{\tau} 
  \int_0^{\tau} j_{ij} \left(
                   \ln\frac{q_{ij}}{q_{ji}}
                   - \ln\frac{\hat{q}_{ij}}{\hat{q}_{ji}} \right).
\end{equation}
The Second Law of Thermodynamics ensures that both $\sigma^{\rm pump}$ and $\sigma^{\rm ss}$ are non-negative on average.
This decomposition is analogous to the decomposition of entropy production used to describe the amount of heat required to maintain a nonequilibrium steady state~\cite{Oono1998,Hatano2001,Esposito2008}.

The excess entropy produced by the stochastic pump, $\sigma^{\rm ex}$ is also non-negative.
The inequality,
\begin{equation}
\frac{1}{\tau} \int_0^{\tau} j_{ij} \ln\frac{q_{ij}}{q_{ji}} \geq
           \hat{j}_{ij} \ln\frac{\hat{q}_{ij}}{\hat{q}_{ji}}.
\end{equation}
follows directly from Jensen's inequality, because $q_{ij}(t)>0$ and $x \ln x$ is a convex function~\cite{Cover2006}.
In the adiabatic limit, the system remains in the instantaneous equilibrium distribution and $\sigma^{\rm ex}$ vanishes.
In this limiting case, $\mathcal{S}_{\avg{\sigma}} = \mathcal{S}_{\avg{\omega}}.$
That is, for slow driving, entropy is only produced in the long time limit if probability is pumped through the network on average. 

In a stochastic pump, the hopping statistics along each edge need not be Poissonian, even in the adiabatic limit~\cite{Sinitsyn2007berry}.
Therefore, the instantaneous dynamics of the nonequilibrium steady state, for which all transitions are purely Poissonian, may not perfectly recapitulate the behavior of the pump.
As further discussed in the Supporting Information, an effective dynamics can be constructed by a periodic solution via Floquet Theory.
In the limit that time-periodic perturbations to the hopping rates are small, the nonequilibrium steady state representation describes the dynamics of the pump at all times.
Numerical simulations using the kinetic Monte Carlo technique provide additional support that this correspondence is robust, as shown in the Supporting Information, and emphasize that statistics converge to the large deviation form on timescales that can easily be accessed in simulations and experiments. 

The mapping determined by the choice~\eqref{eq:sflow} yields a universal bound on current fluctuations in weakly driven stochastic pumps, akin to the thermodynamic uncertainty relations recently discovered for nonequilibrium steady states~\cite{Barato2015,Pietzonka2016,Pietzonka2016universal,Gingrich2016,Gingrich2016jpa}. 
In the perturbative limit, the rate function for any generalized current $j$ is subject to a quadratic bound determined by the steady state entropy production rate, 
\begin{equation}
  I^{\rm pump}(j) \leq \frac{(j-\hat{j})^2}{4 \hat{j}^2 / \sigma^{\rm ss}}.
\end{equation}
The bound is maximally tight because incorporating the excess entropy production only reduces the curvature of the quadratic form.
The lack of Poisson statistics for the pump makes it unlikely that the bound holds in full generality, but numerical evidence suggests that it is quite robust.

{\emph{Acknowledgments}---.}
It is a pleasure to thank Hugo Touchette, Todd Gingrich, Suri Vaikuntanathan, and Phillip Geissler for their useful feedback on this work. Funding for this research was provided by the National Science Foundation Graduate Research Fellowship.

\clearpage
\setcounter{equation}{0}
\setcounter{figure}{0}
\onecolumngrid
\renewcommand{\theequation}{S\arabic{equation}}
\renewcommand{\thefigure}{S\arabic{figure}}
\renewcommand{\bibnumfmt}[1]{[S#1]}
\renewcommand{\citenumfont}[1]{S#1}
\section*{Supporting Information}
\setcounter{page}{1}
\appendix
\section{Monte Carlo Sampling}

To probe the rate of convergence of the large deviation form for current fluctuations in both the pump and the nonequilibrium steady state, we simulated the dynamics using kinetic Monte Carlo (KMC) sampling.
In the case of the nonequilibrium steady state, standard algorithms can be employed~\cite{Bortz1975,Voter2007}.
However, the procedure must be modified slightly to sample the pump, where the rates are time dependent.
To perform the simulations, we follow Ref.~\cite{Prados1997}. 
We note that the probability of escape from state $i$ in time $\Delta t$ is,
\begin{equation}
\exp \left( -\int_{t}^{t+\Delta t} W_{ii}(t) dt \right).
\end{equation}
Thus, we choose a random number $r\in (0,1]$ and compute $\Delta t$ by numerically solving the following equation,
\begin{equation}
\ln(r) = -\int_{t}^{t+\Delta t} W_{ii}(t) dt.
\end{equation}
Once $\Delta t$ is determined, a new state is selected in proportion to the flow from the current state into the new state. 
Consider an ordered list of the rates at time $t+\Delta t$, i.e., $\{ W_{ji} \}_{j\neq i}$ with $W_{ji} < W_{(j+1)i}$ for all $j$.
We define
\begin{equation}
R_{ji}(t+\Delta t) = \sum_{k=1}^j W_{ki}(t+\Delta t).
\end{equation}
Next, we choose a random number $r'\in (0,1]$ and perform a binary search to determine $j$ such that,
\begin{equation}
  R_{ji}(t+\Delta t) \leq r' W_{ii}(t+\Delta t) \leq R_{(j+1)i}(t+\Delta t).
\end{equation}

\begin{figure}[h!]
\includegraphics[width=\linewidth]{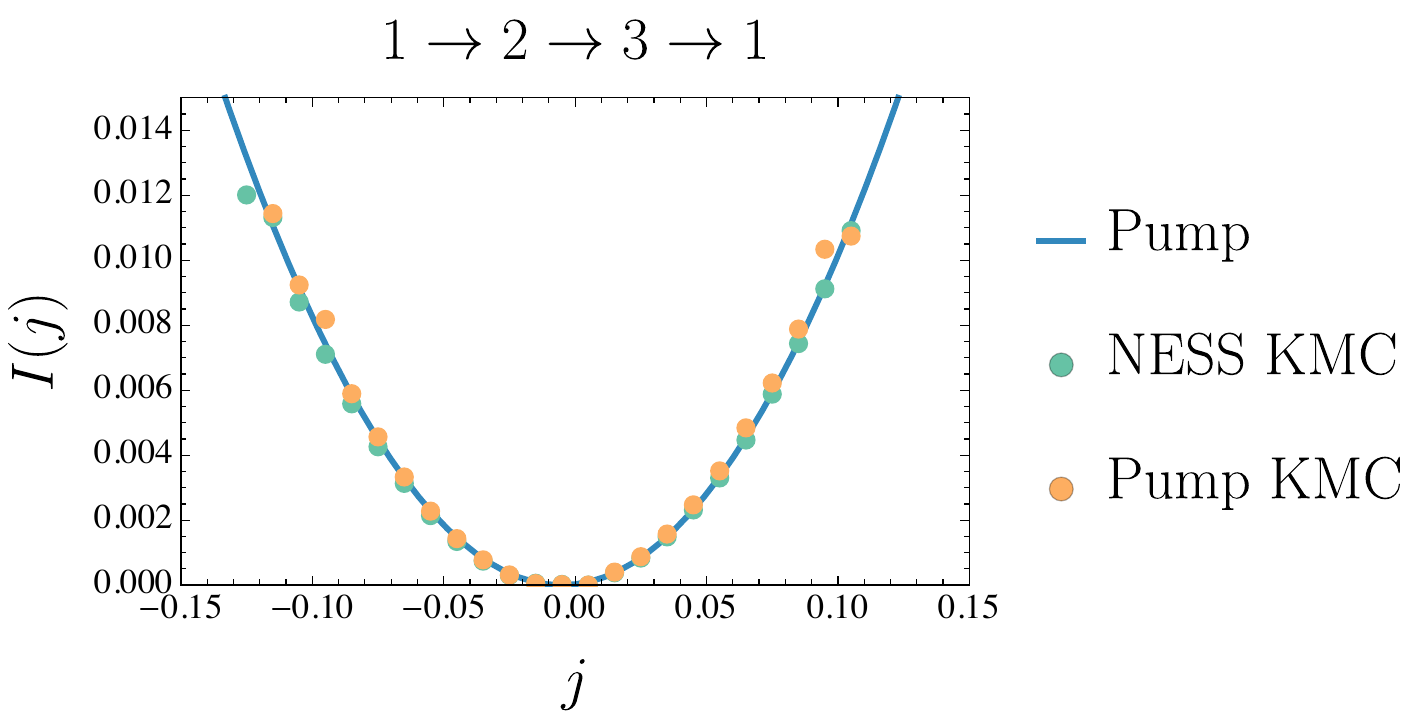}
\caption{Results from kinetic Monte Carlo sampling show good agreement with the asymptotic limit for the stochastic pump. The nonequilibrium steady state simulations use the rate matrix determined by $\mathcal{S}_{\avg{q}}$. The KMC data for the stochastic pump was sampled with the Monte Carlo procedure described in the Supporting Information.}
\label{fig:sampling}
\end{figure}

We collected $1\times 10^6$ independent trajectories for both the stochastic pump and the nonequilibrium steady state representations with an observation time $t_{\rm obs} = 1000.$
We computed the scalar current $j$ around the cycle $1 \to 2 \to 3 \to 1$ and plotted $-1/t_{\rm obs} \ln p_{\rm sim}(j).$
The results, as shown in Fig.~\ref{fig:sampling}, are in good agreement with the asymptotic form of the rate function.
The modest timescale over which the large deviation form is adopted emphasizes the practical implications of these predictions. 

\appendix
\section{Effective Stationary Process}
The time-periodic master equation can be written, 
\begin{equation}
\begin{aligned}
  \partial_t p_i(t) &= \sum_{j} W_{ij}(t) p_j(t), \\
  p_i(0) &= p^{\rm init}_i.
\end{aligned}
\label{eq:me}
\end{equation}
The rates are assumed to vary periodically in time with period $\tau$ so that $W_{ij}(t + \tau) = W_{ij}(t)$.
This equation admits a formal solution using the time-order exponential operator $\overrightarrow{\rm exp},$
\begin{align}
  p(t + \Delta t) &= \overrightarrow{\rm exp} \left( \int_t^{t+\Delta t}dt'\  W(t')\right) p(t), \\
  &\equiv \mathcal{G}(t,t+\Delta t) p(t).
\end{align}
At long times, the solution becomes periodic, up to an exponential factor called the Floquet multiplier.
In our case, there are no sources or sinks for the probability, so the exponents vanish (cf. Ref.~\cite{Talkner1999}).
The propagator $\mathcal{G}(t,t')$ is itself a periodic function of time by Floquet's theorem.
Further, by the semi-group property, 
\begin{equation}
  \mathcal{G}(0,n\tau) = \mathcal{G}^n(0,\tau) \equiv \mathcal{G}^{n}(\tau).
\end{equation}
In the long time limit, $\mathcal{G}(\tau)$ is discrete time rate matrix that propagates probability through the network. 
By the law of large numbers, the average flow $q_{ij}^{\rm eff}$ along each edge determined by $\mathcal{G}$ must match the average in the periodic steady state $\hat{q}_{ij}.$
This correspondence is exact.

It remains to compare the current fluctuations of the non-homogeneous Markov process with current fluctuations in the effective process.
A correspondence between the nonequilibrium steady state and pump will hold in the long time limit if the deviations from Poisson statistics determined by the periodic averages can be neglected.
We consider the Fourier representation of the periodic dynamics in its periodic steady state; the right hand side of Eq.~\eqref{eq:me} becomes,
\begin{equation}
  \sum_j \sum_{k,l} \tilde{W}_{ij}(k-l) \tilde{p}_j^{\rm ps}(l) e^{2\pi i k t}.
\end{equation}
If the time-periodic perturbation to the hopping rates is small, then we can neglect the higher Fourier coefficients and retain only the $k=0$ contribution, 
\begin{align}
  & \sum_j \tilde{W}_{ij}(0) \sum_{l}  \tilde{p}_j^{\rm ps}(l) e^{2\pi i l t},\\
  &= \sum_j \hat{W}_{ij} p^{\rm ps}_j(t).
\end{align}
Because $\tilde{W}_{ij}(0)$ is the time periodic average, the transition rates are given by the average rate of hopping over the course of the period.
The fluctuations, in this case, will be dominated by the long-time properties of $\mathcal{G}$.
Empirically, the pump rate functions show robust agreement with the nonequilibrium steady state representation for a wide range of different pumping protocols and networks.


\begin{thebibliography}{32}%
\makeatletter
\providecommand \@ifxundefined [1]{%
 \@ifx{#1\undefined}
}%
\providecommand \@ifnum [1]{%
 \ifnum #1\expandafter \@firstoftwo
 \else \expandafter \@secondoftwo
 \fi
}%
\providecommand \@ifx [1]{%
 \ifx #1\expandafter \@firstoftwo
 \else \expandafter \@secondoftwo
 \fi
}%
\providecommand \natexlab [1]{#1}%
\providecommand \enquote  [1]{``#1''}%
\providecommand \bibnamefont  [1]{#1}%
\providecommand \bibfnamefont [1]{#1}%
\providecommand \citenamefont [1]{#1}%
\providecommand \href@noop [0]{\@secondoftwo}%
\providecommand \href [0]{\begingroup \@sanitize@url \@href}%
\providecommand \@href[1]{\@@startlink{#1}\@@href}%
\providecommand \@@href[1]{\endgroup#1\@@endlink}%
\providecommand \@sanitize@url [0]{\catcode `\\12\catcode `\$12\catcode
  `\&12\catcode `\#12\catcode `\^12\catcode `\_12\catcode `\%12\relax}%
\providecommand \@@startlink[1]{}%
\providecommand \@@endlink[0]{}%
\providecommand \url  [0]{\begingroup\@sanitize@url \@url }%
\providecommand \@url [1]{\endgroup\@href {#1}{\urlprefix }}%
\providecommand \urlprefix  [0]{URL }%
\providecommand \Eprint [0]{\href }%
\providecommand \doibase [0]{http://dx.doi.org/}%
\providecommand \selectlanguage [0]{\@gobble}%
\providecommand \bibinfo  [0]{\@secondoftwo}%
\providecommand \bibfield  [0]{\@secondoftwo}%
\providecommand \translation [1]{[#1]}%
\providecommand \BibitemOpen [0]{}%
\providecommand \bibitemStop [0]{}%
\providecommand \bibitemNoStop [0]{.\EOS\space}%
\providecommand \EOS [0]{\spacefactor3000\relax}%
\providecommand \BibitemShut  [1]{\csname bibitem#1\endcsname}%
\let\auto@bib@innerbib\@empty
\bibitem [{\citenamefont {Yoshida}\ \emph {et~al.}(2001)\citenamefont
  {Yoshida}, \citenamefont {Muneyuki},\ and\ \citenamefont
  {Hisabori}}]{Yoshida2001}%
  \BibitemOpen
  \bibfield  {author} {\bibinfo {author} {\bibfnamefont {M.}~\bibnamefont
  {Yoshida}}, \bibinfo {author} {\bibfnamefont {E.}~\bibnamefont {Muneyuki}}, \
  and\ \bibinfo {author} {\bibfnamefont {T.}~\bibnamefont {Hisabori}},\ }\href
  {\doibase 10.1038/35089509} {\bibfield  {journal} {\bibinfo  {journal} {Nat.
  Rev. Mol. Cell Biol.}\ }\textbf {\bibinfo {volume} {2}},\ \bibinfo {pages}
  {669} (\bibinfo {year} {2001})}\BibitemShut {NoStop}%
\bibitem [{\citenamefont {Wilson}\ \emph {et~al.}(2016)\citenamefont {Wilson},
  \citenamefont {Sol{\`a}}, \citenamefont {Carlone}, \citenamefont {Goldup},
  \citenamefont {Lebrasseur},\ and\ \citenamefont {Leigh}}]{Wilson2016}%
  \BibitemOpen
  \bibfield  {author} {\bibinfo {author} {\bibfnamefont {M.~R.}\ \bibnamefont
  {Wilson}}, \bibinfo {author} {\bibfnamefont {J.}~\bibnamefont {Sol{\`a}}},
  \bibinfo {author} {\bibfnamefont {A.}~\bibnamefont {Carlone}}, \bibinfo
  {author} {\bibfnamefont {S.~M.}\ \bibnamefont {Goldup}}, \bibinfo {author}
  {\bibfnamefont {N.}~\bibnamefont {Lebrasseur}}, \ and\ \bibinfo {author}
  {\bibfnamefont {D.~A.}\ \bibnamefont {Leigh}},\ }\href {\doibase 
  10.1038/nature18013} {\bibfield  {journal} {\bibinfo  {journal} {Nature}\
  }\textbf {\bibinfo {volume} {534}},\ \bibinfo {pages} {235} (\bibinfo {year}
  {2016})}\BibitemShut {NoStop}%
\bibitem [{\citenamefont {Browne}\ and\ \citenamefont
  {Feringa}(2006)}]{Browne2006}%
  \BibitemOpen
  \bibfield  {author} {\bibinfo {author} {\bibfnamefont {W.~R.}\ \bibnamefont
  {Browne}}\ and\ \bibinfo {author} {\bibfnamefont {B.~L.}\ \bibnamefont
  {Feringa}},\ }\href {\doibase 10.1038/nnano.2006.45} {\bibfield  {journal}
  {\bibinfo  {journal} {Nat. Nanotechnology}\ }\textbf {\bibinfo {volume}
  {1}},\ \bibinfo {pages} {25} (\bibinfo {year} {2006})}\BibitemShut {NoStop}%
\bibitem [{\citenamefont {Sinitsyn}\ and\ \citenamefont
  {Nemenman}(2007{\natexlab{a}})}]{Sinitsyn2007}%
  \BibitemOpen
  \bibfield  {author} {\bibinfo {author} {\bibfnamefont {N.~A.}\ \bibnamefont
  {Sinitsyn}}\ and\ \bibinfo {author} {\bibfnamefont {I.}~\bibnamefont
  {Nemenman}},\ }\href {\doibase 10.1103/PhysRevLett.99.220408} {\bibfield
  {journal} {\bibinfo  {journal} {Phys. Rev. Lett.}\ }\textbf {\bibinfo
  {volume} {99}},\ \bibinfo {pages} {220408} (\bibinfo {year}
  {2007}{\natexlab{a}})}\BibitemShut {NoStop}%
\bibitem [{\citenamefont {Horowitz}\ and\ \citenamefont
  {Jarzynski}(2009)}]{Horowitz2009}%
  \BibitemOpen
  \bibfield  {author} {\bibinfo {author} {\bibfnamefont {J.~M.}\ \bibnamefont
  {Horowitz}}\ and\ \bibinfo {author} {\bibfnamefont {C.}~\bibnamefont
  {Jarzynski}},\ }\href {\doibase 10.1007/s10955-009-9818-x} {\bibfield
  {journal} {\bibinfo  {journal} {J. Stat. Phys.}\ }\textbf {\bibinfo {volume}
  {136}},\ \bibinfo {pages} {917} (\bibinfo {year} {2009})}\BibitemShut
  {NoStop}%
\bibitem [{\citenamefont {Sinitsyn}(2009)}]{Sinitsyn2009}%
  \BibitemOpen
  \bibfield  {author} {\bibinfo {author} {\bibfnamefont {N.}~\bibnamefont
  {Sinitsyn}},\ }\href {\doibase 10.1088/1751-8113/42/19/193001} {\bibfield
  {journal} {\bibinfo  {journal} {J. Phys. A}\ }\textbf {\bibinfo {volume}
  {42}},\ \bibinfo {pages} {193001} (\bibinfo {year} {2009})}\BibitemShut
  {NoStop}%
\bibitem [{\citenamefont {Mart\'{i}nez}\ \emph {et~al.}(2016)\citenamefont
  {Mart\'{i}nez}, \citenamefont {Rold\'{a}n}, \citenamefont {Dinis},
  \citenamefont {Petrov}, \citenamefont {Parrondo},\ and\ \citenamefont
  {Rica}}]{Martinez2016}%
  \BibitemOpen
  \bibfield  {author} {\bibinfo {author} {\bibfnamefont {I.~A.}\ \bibnamefont
  {Mart\'{i}nez}}, \bibinfo {author} {\bibfnamefont {E.}~\bibnamefont
  {Rold\'{a}n}}, \bibinfo {author} {\bibfnamefont {L.}~\bibnamefont {Dinis}},
  \bibinfo {author} {\bibfnamefont {D.}~\bibnamefont {Petrov}}, \bibinfo
  {author} {\bibfnamefont {J.~M.~R.}\ \bibnamefont {Parrondo}}, \ and\ \bibinfo
  {author} {\bibfnamefont {R.~A.}\ \bibnamefont {Rica}},\ }\href {\doibase
  10.1038/nphys3518} {\bibfield  {journal} {\bibinfo  {journal} {Nat. Phys.}\
  }\textbf {\bibinfo {volume} {12}},\ \bibinfo {pages} {67} (\bibinfo {year}
  {2016})}\BibitemShut {NoStop}%
\bibitem [{\citenamefont {Rahav}\ \emph {et~al.}(2008)\citenamefont {Rahav},
  \citenamefont {Horowitz},\ and\ \citenamefont {Jarzynski}}]{Rahav2008}%
  \BibitemOpen
  \bibfield  {author} {\bibinfo {author} {\bibfnamefont {S.}~\bibnamefont
  {Rahav}}, \bibinfo {author} {\bibfnamefont {J.}~\bibnamefont {Horowitz}}, \
  and\ \bibinfo {author} {\bibfnamefont {C.}~\bibnamefont {Jarzynski}},\ }\href
  {\doibase 10.1103/PhysRevLett.101.140602} {\bibfield  {journal} {\bibinfo
  {journal} {Phys. Rev. Lett.}\ }\textbf {\bibinfo {volume} {101}},\ \bibinfo
  {pages} {140602} (\bibinfo {year} {2008})}\BibitemShut {NoStop}%
\bibitem [{\citenamefont {Mandal}\ and\ \citenamefont
  {Jarzynski}(2011)}]{Mandal2011}%
  \BibitemOpen
  \bibfield  {author} {\bibinfo {author} {\bibfnamefont {D.}~\bibnamefont
  {Mandal}}\ and\ \bibinfo {author} {\bibfnamefont {C.}~\bibnamefont
  {Jarzynski}},\ }\href {\doibase 10.1088/1742-5468/2011/10/P10006} {\bibfield
  {journal} {\bibinfo  {journal} {J. Stat. Mech. Theor. Exp}\ }\textbf
  {\bibinfo {volume} {2011}},\ \bibinfo {pages} {P10006} (\bibinfo {year}
  {2011})}\BibitemShut {NoStop}%
\bibitem [{\citenamefont {Raz}\ \emph {et~al.}(2016)\citenamefont {Raz},
  \citenamefont {Suba\c{s}i},\ and\ \citenamefont {Jarzynski}}]{Raz2016}%
  \BibitemOpen
  \bibfield  {author} {\bibinfo {author} {\bibfnamefont {O.}~\bibnamefont
  {Raz}}, \bibinfo {author} {\bibfnamefont {Y.}~\bibnamefont {Suba\c{s}i}}, \
  and\ \bibinfo {author} {\bibfnamefont {C.}~\bibnamefont {Jarzynski}},\ }\href
  {\doibase 10.1103/PhysRevX.6.021022} {\bibfield  {journal} {\bibinfo
  {journal} {Phys. Rev. X}\ }\textbf {\bibinfo {volume} {6}},\ \bibinfo {pages}
  {021022} (\bibinfo {year} {2016})}\BibitemShut {NoStop}%
\bibitem [{\citenamefont {Verley}\ \emph {et~al.}(2014)\citenamefont {Verley},
  \citenamefont {Esposito}, \citenamefont {Willaert},\ and\ \citenamefont
  {Van~den Broeck}}]{Verley2014}%
  \BibitemOpen
  \bibfield  {author} {\bibinfo {author} {\bibfnamefont {G.}~\bibnamefont
  {Verley}}, \bibinfo {author} {\bibfnamefont {M.}~\bibnamefont {Esposito}},
  \bibinfo {author} {\bibfnamefont {T.}~\bibnamefont {Willaert}}, \ and\
  \bibinfo {author} {\bibfnamefont {C.}~\bibnamefont {Van~den Broeck}},\ }\href
  {\doibase 10.1038/ncomms5721} {\bibfield  {journal} {\bibinfo  {journal}
  {Nat. Commun.}\ }\textbf {\bibinfo {volume} {5}} (\bibinfo {year} {2014}),\
  10.1038/ncomms5721}\BibitemShut {NoStop}%
\bibitem [{\citenamefont {Gingrich}\ \emph {et~al.}(2014)\citenamefont
  {Gingrich}, \citenamefont {Rotskoff}, \citenamefont {Vaikuntanathan},\ and\
  \citenamefont {Geissler}}]{Gingrich2014}%
  \BibitemOpen
  \bibfield  {author} {\bibinfo {author} {\bibfnamefont {T.~R.}\ \bibnamefont
  {Gingrich}}, \bibinfo {author} {\bibfnamefont {G.~M.}\ \bibnamefont
  {Rotskoff}}, \bibinfo {author} {\bibfnamefont {S.}~\bibnamefont
  {Vaikuntanathan}}, \ and\ \bibinfo {author} {\bibfnamefont {P.~L.}\
  \bibnamefont {Geissler}},\ }\href {\doibase 10.1088/1367-2630/16/10/102003}
  {\bibfield  {journal} {\bibinfo  {journal} {New J. Phys.}\ }\textbf {\bibinfo
  {volume} {16}},\ \bibinfo {pages} {102003} (\bibinfo {year}
  {2014})}\BibitemShut {NoStop}%
\bibitem [{\citenamefont {Zia}\ and\ \citenamefont
  {Schmittmann}(2007)}]{Zia2007}%
  \BibitemOpen
  \bibfield  {author} {\bibinfo {author} {\bibfnamefont {R.}~\bibnamefont
  {Zia}}\ and\ \bibinfo {author} {\bibfnamefont {B.}~\bibnamefont
  {Schmittmann}},\ }\href {\doibase 10.1088/1742-5468/2007/07/P07012}
  {\bibfield  {journal} {\bibinfo  {journal} {J. Stat. Mech. Theor. Exp}\
  }\textbf {\bibinfo {volume} {2007}},\ \bibinfo {pages} {P07012} (\bibinfo
  {year} {2007})}\BibitemShut {NoStop}%
\bibitem [{\citenamefont {Maes}\ and\ \citenamefont
  {Neto{\v{c}}n{\`y}}(2008)}]{Maes2008}%
  \BibitemOpen
  \bibfield  {author} {\bibinfo {author} {\bibfnamefont {C.}~\bibnamefont
  {Maes}}\ and\ \bibinfo {author} {\bibfnamefont {K.}~\bibnamefont
  {Neto{\v{c}}n{\`y}}},\ }\href {\doibase 10.1209/0295-5075/82/30003}
  {\bibfield  {journal} {\bibinfo  {journal} {Euro. Phys. Lett.}\ }\textbf
  {\bibinfo {volume} {82}},\ \bibinfo {pages} {30003} (\bibinfo {year}
  {2008})}\BibitemShut {NoStop}%
\bibitem [{\citenamefont {Bertini}\ \emph {et~al.}(2015)\citenamefont
  {Bertini}, \citenamefont {Faggionato},\ and\ \citenamefont
  {Gabrielli}}]{Bertini2015Large}%
  \BibitemOpen
  \bibfield  {author} {\bibinfo {author} {\bibfnamefont {L.}~\bibnamefont
  {Bertini}}, \bibinfo {author} {\bibfnamefont {A.}~\bibnamefont {Faggionato}},
  \ and\ \bibinfo {author} {\bibfnamefont {D.}~\bibnamefont {Gabrielli}},\
  }\href {\doibase 10.1214/14-AIHP601} {\bibfield  {journal} {\bibinfo
  {journal} {Ann. I. H. Poincar{\'e} -- PR}\ }\textbf {\bibinfo {volume}
  {51}},\ \bibinfo {pages} {867} (\bibinfo {year} {2015})}\BibitemShut
  {NoStop}%
\bibitem [{\citenamefont {Barato}\ and\ \citenamefont
  {Chetrite}(2015)}]{Barato2015Formal}%
  \BibitemOpen
  \bibfield  {author} {\bibinfo {author} {\bibfnamefont {A.~C.}\ \bibnamefont
  {Barato}}\ and\ \bibinfo {author} {\bibfnamefont {R.}~\bibnamefont
  {Chetrite}},\ }\href {\doibase 10.1007/s10955-015-1283-0} {\bibfield
  {journal} {\bibinfo  {journal} {J. Stat. Phys.}\ }\textbf {\bibinfo {volume}
  {160}},\ \bibinfo {pages} {1154} (\bibinfo {year} {2015})}\BibitemShut
  {NoStop}%
\bibitem [{\citenamefont {Gingrich}\ \emph
  {et~al.}(2016{\natexlab{a}})\citenamefont {Gingrich}, \citenamefont
  {Rotskoff},\ and\ \citenamefont {Horowitz}}]{Gingrich2016jpa}%
  \BibitemOpen
  \bibfield  {author} {\bibinfo {author} {\bibfnamefont {T.~R.}\ \bibnamefont
  {Gingrich}}, \bibinfo {author} {\bibfnamefont {G.~M.}\ \bibnamefont
  {Rotskoff}}, \ and\ \bibinfo {author} {\bibfnamefont {J.~M.}\ \bibnamefont
  {Horowitz}},\ }\href@noop {} {\  (\bibinfo {year} {2016}{\natexlab{a}})},\
  \Eprint {http://arxiv.org/abs/1609.07131v1} {arXiv:1609.07131v1
  [cond-mat.stat-mech]} \BibitemShut {NoStop}%
\bibitem [{\citenamefont {Touchette}(2009)}]{Touchette2009}%
  \BibitemOpen
  \bibfield  {author} {\bibinfo {author} {\bibfnamefont {H.}~\bibnamefont
  {Touchette}},\ }\href {\doibase 10.1016/j.physrep.2009.05.002} {\bibfield
  {journal} {\bibinfo  {journal} {Phys. Rep.}\ }\textbf {\bibinfo {volume}
  {478}},\ \bibinfo {pages} {1} (\bibinfo {year} {2009})}\BibitemShut {NoStop}%
\bibitem [{\citenamefont {Oono}\ and\ \citenamefont
  {Paniconi}(1998)}]{Oono1998}%
  \BibitemOpen
  \bibfield  {author} {\bibinfo {author} {\bibfnamefont {Y.}~\bibnamefont
  {Oono}}\ and\ \bibinfo {author} {\bibfnamefont {M.}~\bibnamefont
  {Paniconi}},\ }\href {\doibase 10.1143/PTPS.130.29} {\bibfield  {journal}
  {\bibinfo  {journal} {Prog. Theor. Phys. Supplement}\ }\textbf {\bibinfo
  {volume} {130}},\ \bibinfo {pages} {29} (\bibinfo {year} {1998})}\BibitemShut
  {NoStop}%
\bibitem [{\citenamefont {Hatano}\ and\ \citenamefont
  {Sasa}(2001)}]{Hatano2001}%
  \BibitemOpen
  \bibfield  {author} {\bibinfo {author} {\bibfnamefont {T.}~\bibnamefont
  {Hatano}}\ and\ \bibinfo {author} {\bibfnamefont {S.-i.}\ \bibnamefont
  {Sasa}},\ }\href {\doibase 10.1103/PhysRevLett.86.3463} {\bibfield  {journal}
  {\bibinfo  {journal} {Phys. Rev. Lett.}\ }\textbf {\bibinfo {volume} {86}},\
  \bibinfo {pages} {3463} (\bibinfo {year} {2001})}\BibitemShut {NoStop}%
\bibitem [{\citenamefont {Esposito}\ and\ \citenamefont {Van~den
  Broeck}(2010{\natexlab{a}})}]{Esposito2008}%
  \BibitemOpen
  \bibfield  {author} {\bibinfo {author} {\bibfnamefont {M.}~\bibnamefont
  {Esposito}}\ and\ \bibinfo {author} {\bibfnamefont {C.}~\bibnamefont {Van~den
  Broeck}},\ }\href {\doibase 10.1103/PhysRevE.82.011143} {\bibfield  {journal}
  {\bibinfo  {journal} {Phys. Rev. E}\ }\textbf {\bibinfo {volume} {82}},\
  \bibinfo {pages} {011143} (\bibinfo {year} {2010}{\natexlab{a}})}\BibitemShut
  {NoStop}%
\bibitem [{\citenamefont {Barato}\ and\ \citenamefont
  {Seifert}(2015)}]{Barato2015}%
  \BibitemOpen
  \bibfield  {author} {\bibinfo {author} {\bibfnamefont {A.~C.}\ \bibnamefont
  {Barato}}\ and\ \bibinfo {author} {\bibfnamefont {U.}~\bibnamefont
  {Seifert}},\ }\href {\doibase 10.1103/PhysRevLett.114.158101} {\bibfield
  {journal} {\bibinfo  {journal} {Phys. Rev. Lett.}\ }\textbf {\bibinfo
  {volume} {114}},\ \bibinfo {pages} {158101} (\bibinfo {year}
  {2015})}\BibitemShut {NoStop}%
\bibitem [{\citenamefont {Pietzonka}\ \emph
  {et~al.}(2016{\natexlab{a}})\citenamefont {Pietzonka}, \citenamefont
  {Barato},\ and\ \citenamefont {Seifert}}]{Pietzonka2016}%
  \BibitemOpen
  \bibfield  {author} {\bibinfo {author} {\bibfnamefont {P.}~\bibnamefont
  {Pietzonka}}, \bibinfo {author} {\bibfnamefont {A.~C.}\ \bibnamefont
  {Barato}}, \ and\ \bibinfo {author} {\bibfnamefont {U.}~\bibnamefont
  {Seifert}},\ }\href {\doibase 10.1088/1751-8113/49/34/34LT01} {\bibfield
  {journal} {\bibinfo  {journal} {J. Phys. A}\ }\textbf {\bibinfo {volume}
  {49}},\ \bibinfo {pages} {34LT01} (\bibinfo {year}
  {2016}{\natexlab{a}})}\BibitemShut {NoStop}%
\bibitem [{\citenamefont {Gingrich}\ \emph
  {et~al.}(2016{\natexlab{b}})\citenamefont {Gingrich}, \citenamefont
  {Horowitz}, \citenamefont {Perunov},\ and\ \citenamefont
  {England}}]{Gingrich2016}%
  \BibitemOpen
  \bibfield  {author} {\bibinfo {author} {\bibfnamefont {T.~R.}\ \bibnamefont
  {Gingrich}}, \bibinfo {author} {\bibfnamefont {J.~M.}\ \bibnamefont
  {Horowitz}}, \bibinfo {author} {\bibfnamefont {N.}~\bibnamefont {Perunov}}, \
  and\ \bibinfo {author} {\bibfnamefont {J.~L.}\ \bibnamefont {England}},\
  }\href {\doibase 10.1103/PhysRevLett.116.120601} {\bibfield  {journal}
  {\bibinfo  {journal} {Phys. Rev. Lett.}\ }\textbf {\bibinfo {volume} {116}},\
  \bibinfo {pages} {120601} (\bibinfo {year} {2016}{\natexlab{b}})}\BibitemShut
  {NoStop}%
\bibitem [{\citenamefont {Lecomte}\ \emph {et~al.}(2007)\citenamefont
  {Lecomte}, \citenamefont {Appert-Rolland},\ and\ \citenamefont {van
  Wijland}}]{Lecomte2007}%
  \BibitemOpen
  \bibfield  {author} {\bibinfo {author} {\bibfnamefont {V.}~\bibnamefont
  {Lecomte}}, \bibinfo {author} {\bibfnamefont {C.}~\bibnamefont
  {Appert-Rolland}}, \ and\ \bibinfo {author} {\bibfnamefont {F.}~\bibnamefont
  {van Wijland}},\ }\href {\doibase 10.1007/s10955-006-9254-0} {\bibfield
  {journal} {\bibinfo  {journal} {J. Stat. Phys.}\ }\textbf {\bibinfo {volume}
  {127}},\ \bibinfo {pages} {51} (\bibinfo {year} {2007})}\BibitemShut
  {NoStop}%
\bibitem [{\citenamefont {Baiesi}\ \emph {et~al.}(2009)\citenamefont {Baiesi},
  \citenamefont {Maes},\ and\ \citenamefont {Wynants}}]{Baiesi2009}%
  \BibitemOpen
  \bibfield  {author} {\bibinfo {author} {\bibfnamefont {M.}~\bibnamefont
  {Baiesi}}, \bibinfo {author} {\bibfnamefont {C.}~\bibnamefont {Maes}}, \ and\
  \bibinfo {author} {\bibfnamefont {B.}~\bibnamefont {Wynants}},\ }\href
  {\doibase 10.1103/PhysRevLett.103.010602} {\bibfield  {journal} {\bibinfo
  {journal} {Phys. Rev. Lett.}\ }\textbf {\bibinfo {volume} {103}},\ \bibinfo
  {pages} {010602} (\bibinfo {year} {2009})}\BibitemShut {NoStop}%
\bibitem [{\citenamefont {Lebowitz}\ and\ \citenamefont
  {Spohn}(1999)}]{Lebowitz1999}%
  \BibitemOpen
  \bibfield  {author} {\bibinfo {author} {\bibfnamefont {J.~L.}\ \bibnamefont
  {Lebowitz}}\ and\ \bibinfo {author} {\bibfnamefont {H.}~\bibnamefont
  {Spohn}},\ }\href {\doibase 10.1023/A:1004589714161} {\bibfield  {journal}
  {\bibinfo  {journal} {J. Stat. Phys.}\ }\textbf {\bibinfo {volume} {95}},\
  \bibinfo {pages} {333} (\bibinfo {year} {1999})}\BibitemShut {NoStop}%
\bibitem [{\citenamefont {Esposito}\ and\ \citenamefont {Van~den
  Broeck}(2010{\natexlab{b}})}]{Esposito2010}%
  \BibitemOpen
  \bibfield  {author} {\bibinfo {author} {\bibfnamefont {M.}~\bibnamefont
  {Esposito}}\ and\ \bibinfo {author} {\bibfnamefont {C.}~\bibnamefont {Van~den
  Broeck}},\ }\href {\doibase 10.1103/PhysRevE.82.011143} {\bibfield  {journal}
  {\bibinfo  {journal} {Phys. Rev. E}\ }\textbf {\bibinfo {volume} {82}},\
  \bibinfo {pages} {011143} (\bibinfo {year} {2010}{\natexlab{b}})}\BibitemShut
  {NoStop}%
\bibitem [{\citenamefont {Cover}\ and\ \citenamefont
  {Thomas}(2006)}]{Cover2006}%
  \BibitemOpen
  \bibfield  {author} {\bibinfo {author} {\bibfnamefont {T.~M.}\ \bibnamefont
  {Cover}}\ and\ \bibinfo {author} {\bibfnamefont {J.~A.}\ \bibnamefont
  {Thomas}},\ }\href@noop {} {\emph {\bibinfo {title} {Elements of information
  theory}}},\ \bibinfo {edition} {2nd}\ ed.\ (\bibinfo  {publisher} {A. John
  Wiley \& Sons},\ \bibinfo {address} {Hoboken, NJ},\ \bibinfo {year}
  {2006})\BibitemShut {NoStop}%
\bibitem [{\citenamefont {Sinitsyn}\ and\ \citenamefont
  {Nemenman}(2007{\natexlab{b}})}]{Sinitsyn2007berry}%
  \BibitemOpen
  \bibfield  {author} {\bibinfo {author} {\bibfnamefont {N.}~\bibnamefont
  {Sinitsyn}}\ and\ \bibinfo {author} {\bibfnamefont {I.}~\bibnamefont
  {Nemenman}},\ }\href {\doibase 10.1209/0295-5075/77/58001} {\bibfield
  {journal} {\bibinfo  {journal} {Euro. Phys. Lett.}\ }\textbf {\bibinfo
  {volume} {77}},\ \bibinfo {pages} {58001} (\bibinfo {year}
  {2007}{\natexlab{b}})}\BibitemShut {NoStop}%
\bibitem [{\citenamefont {Pietzonka}\ \emph
  {et~al.}(2016{\natexlab{b}})\citenamefont {Pietzonka}, \citenamefont
  {Barato},\ and\ \citenamefont {Seifert}}]{Pietzonka2016universal}%
  \BibitemOpen
  \bibfield  {author} {\bibinfo {author} {\bibfnamefont {P.}~\bibnamefont
  {Pietzonka}}, \bibinfo {author} {\bibfnamefont {A.~C.}\ \bibnamefont
  {Barato}}, \ and\ \bibinfo {author} {\bibfnamefont {U.}~\bibnamefont
  {Seifert}},\ }\href {\doibase 10.1103/PhysRevE.93.052145} {\bibfield
  {journal} {\bibinfo  {journal} {Phys. Rev. E}\ }\textbf {\bibinfo {volume}
  {93}},\ \bibinfo {pages} {052145} (\bibinfo {year}
  {2016}{\natexlab{b}})}\BibitemShut {NoStop}%
\end{thebibliography}

\begin{thebibliography}{3}%
\makeatletter
\providecommand \@ifxundefined [1]{%
 \@ifx{#1\undefined}
}%
\providecommand \@ifnum [1]{%
 \ifnum #1\expandafter \@firstoftwo
 \else \expandafter \@secondoftwo
 \fi
}%
\providecommand \@ifx [1]{%
 \ifx #1\expandafter \@firstoftwo
 \else \expandafter \@secondoftwo
 \fi
}%
\providecommand \natexlab [1]{#1}%
\providecommand \enquote  [1]{``#1''}%
\providecommand \bibnamefont  [1]{#1}%
\providecommand \bibfnamefont [1]{#1}%
\providecommand \citenamefont [1]{#1}%
\providecommand \href@noop [0]{\@secondoftwo}%
\providecommand \href [0]{\begingroup \@sanitize@url \@href}%
\providecommand \@href[1]{\@@startlink{#1}\@@href}%
\providecommand \@@href[1]{\endgroup#1\@@endlink}%
\providecommand \@sanitize@url [0]{\catcode `\\12\catcode `\$12\catcode
  `\&12\catcode `\#12\catcode `\^12\catcode `\_12\catcode `\%12\relax}%
\providecommand \@@startlink[1]{}%
\providecommand \@@endlink[0]{}%
\providecommand \url  [0]{\begingroup\@sanitize@url \@url }%
\providecommand \@url [1]{\endgroup\@href {#1}{\urlprefix }}%
\providecommand \urlprefix  [0]{URL }%
\providecommand \Eprint [0]{\href }%
\providecommand \doibase [0]{http://dx.doi.org/}%
\providecommand \selectlanguage [0]{\@gobble}%
\providecommand \bibinfo  [0]{\@secondoftwo}%
\providecommand \bibfield  [0]{\@secondoftwo}%
\providecommand \translation [1]{[#1]}%
\providecommand \BibitemOpen [0]{}%
\providecommand \bibitemStop [0]{}%
\providecommand \bibitemNoStop [0]{.\EOS\space}%
\providecommand \EOS [0]{\spacefactor3000\relax}%
\providecommand \BibitemShut  [1]{\csname bibitem#1\endcsname}%
\let\auto@bib@innerbib\@empty
\bibitem [{\citenamefont {Bortz}\ \emph {et~al.}(1975)\citenamefont {Bortz},
  \citenamefont {Kalos},\ and\ \citenamefont {Lebowitz}}]{Bortz1975}%
  \BibitemOpen
  \bibfield  {author} {\bibinfo {author} {\bibfnamefont {A.~B.}\ \bibnamefont
  {Bortz}}, \bibinfo {author} {\bibfnamefont {M.~H.}\ \bibnamefont {Kalos}}, \
  and\ \bibinfo {author} {\bibfnamefont {J.~L.}\ \bibnamefont {Lebowitz}},\
  }\href {\doibase 10.1016/0021-9991(75)90060-1} {\bibfield  {journal}
  {\bibinfo  {journal} {J. Comput. Phys.}\ }\textbf {\bibinfo {volume} {17}},\
  \bibinfo {pages} {10} (\bibinfo {year} {1975})}\BibitemShut {NoStop}%
\bibitem [{\citenamefont {Voter}(2007)}]{Voter2007}%
  \BibitemOpen
  \bibfield  {author} {\bibinfo {author} {\bibfnamefont {A.~F.}\ \bibnamefont
  {Voter}},\ }in\ \href@noop {} {\emph {\bibinfo {booktitle} {Radiation Effects
  in Solids}}}\ (\bibinfo  {publisher} {Springer, Netherlands},\ \bibinfo
  {year} {2007})\ pp.\ \bibinfo {pages} {1--23}\BibitemShut {NoStop}%
\bibitem [{\citenamefont {Prados}\ \emph {et~al.}(1997)\citenamefont {Prados},
  \citenamefont {Brey},\ and\ \citenamefont {S{\'a}nchez-Rey}}]{Prados1997}%
  \BibitemOpen
  \bibfield  {author} {\bibinfo {author} {\bibfnamefont {A.}~\bibnamefont
  {Prados}}, \bibinfo {author} {\bibfnamefont {J.~J.}\ \bibnamefont {Brey}}, \
  and\ \bibinfo {author} {\bibfnamefont {B.}~\bibnamefont {S{\'a}nchez-Rey}},\
  }\href {\doibase 10.1007/BF02765541} {\bibfield  {journal} {\bibinfo
  {journal} {J. Stat. Phys.}\ }\textbf {\bibinfo {volume} {89}},\ \bibinfo
  {pages} {709} (\bibinfo {year} {1997})}\BibitemShut {NoStop}%
\bibitem [{\citenamefont {Talkner}(1999)}]{Talkner1999}%
  \BibitemOpen
  \bibfield  {author} {\bibinfo {author} {\bibfnamefont {P.}~\bibnamefont
  {Talkner}},\ }\href {\doibase 10.1088/1367-2630/1/1/004} {\bibfield  {journal}
  {\bibinfo  {journal} {New J. Phys.}\ }\textbf {\bibinfo {volume} {1}},\
  \bibinfo {pages} {4} (\bibinfo {year} {1999})}\BibitemShut {NoStop}%
\end{thebibliography}
\end{document}